# Raman signature of multiple phase transitions and quasi-particle excitations in putative Kitaev spin liquid candidate $Na_2Co_2TeO_6$

Atul G. Chakkar[1,*], Chaitanya B. Auti[1], Deepu Kumar[1], Nirmalya Jana[2], Koushik Pal[2], and Pradeep Kumar[1,#]

[1] School of Physical Sciences, Indian Institute of Technology Mandi, Mandi-175005, India

[2] Department of Physics, Indian Institute of Technology Kanpur, Kanpur-208016, India

**Abstract:**

Two-dimensional cobalt-based honeycomb oxide $Na_2Co_2TeO_6$ is an important candidate for the realization of Kitaev physics and may provide future platform for the quantum computation and quantum technology. Here, we report an in-depth temperature as well as polarization dependent inelastic light scattering (Raman) measurements on the single crystals of quasi-two-dimensional $Na_2Co_2TeO_6$. Our study reveal signature of multiple phase transitions i.e. long-range zigzag antiferromagnetic transition ($T_N$) at ~ 30 K, ferroelectric transition ($T_{FE}$) at ~ 70 K, and a crossover from pure paramagnetic phase to a quantum paramagnetic phase around ~ 150 K reflected in the renormalized self-energy parameters of the Raman active phonon modes. A distinct signature of spin reorientation deep into the AFM phase around $T_{SR}$ ~ 17 K is observed, marked by the clear change in the frequency and linewidth slopes. We also observed an asymmetric phonon mode in the low frequency region, and it appears below the transition temperature ~ 50 K, attributed to the magnetic excitations other than the magnon. The Raman signature of multiple crystal-field excitations at low temperature along with lifting of the Kramer's degeneracy is also observed. Signature of the underlying broad magnetic continuum in the quantum paramagnetic phase and its temperature dependence suggest presence of frustrated magnetic interaction in the quantum paramagnetic phase below ~ 150 K.

[*]E-mail: atulchakkar16@gmail.com
[#]E-mail: pkumar@iitmandi.ac.in

**Introduction:**

Recent research on strongly correlated materials, in particular understanding of highly entangled phases such as Quantum Spin Liquid (QSL), is driven by their potential applications in quantum technology and quantum computation due to their exotic properties such as fractionalization of the elementary excitations, topological order, and emergent gauge fields [1–3]. According to the Mermin-Wagner theorem, magnetic anisotropy balances thermal fluctuations that affect long-range magnetic order in two-dimensional (2D) Van der Waal's systems [4,5]. Unlike conventional symmetry broken state of matter, the ground state wavefunction in QSL is characterized by the long-range quantum entanglement among the local degrees of freedom. In these materials, long-range magnetic ordering cannot form due to strong quantum fluctuations resulting from microscopic degeneracy. Kitaev model on a 2D honeycomb lattice with spin S=1/2 is exactly solvable and the spins fractionalize into Majorana fermions, forming a topological QSL in the ground state [6]. For the realization of Kitaev interactions, two essential prerequisites are (1) the magnetic ions having spin-orbit entangled Kramer's degenerate ground state and (2) these ions be arranged on a 2D honeycomb lattice structure [6–9]. The Hamiltonian for Kitaev-Heisenberg model is given as $H = \sum_{\langle i,j \rangle_\gamma} K_\gamma S_i^\gamma S_j^\gamma + \sum J(\vec{S}_i \cdot \vec{S}_j)$, where K is the Kitaev coupling constant, $J$ is the Heinsberg exchange parameter, and $\gamma = x, y, z$ [10].

An extensively studied 2D Kitaev QSL candidates α-RuCl$_3$, H$_3$LiIr$_2$O$_6$, Cu$_3$LiIr$_2$O$_6$, Ag$_3$LiIr$_2$O$_6$, Na$_2$IrO$_3$, and α-Li$_2$IrO$_3$ shows the exotic properties and rich physics [11–19]. To explore the Kitaev physics, the original Jackeli and Khaliullin mechanism has been extended to a high spin $d^7$ electron configuration [10,20–22] for the possible realization of a Kitaev physics. This open doors for exploring new materials with new possibilities in fundamental as well as applicative research fields. Current progress in this field suggests spin-orbit coupled 2D honeycomb magnets made of cobalt are promising candidate for the realization of the Kitaev model [10,20,21]. The recently reported materials i.e., $Li_3Co_2SbO_6$, $Na_3Co_2SbO_6$, and $Na_2Co_2TeO_6$ in high spin $d^7$ configuration shows interesting properties gravitating towards QSL physics [10,21,23,24]. Among which $Na_2Co_2TeO_6$ (NCTO), a *3d* cobalt-based honeycomb lattice which is a putative Kitaev candidate is gaining increased attention for its unique properties and broad range of applications. Ground state (GS) of NCTO is a zigzag anti-ferromagnetic (AFM) with entangled spin and orbital degrees of freedoms and such a zigzag magnetic GS was also reported for other honeycomb lattice such as α-$RuCl_3$ [25] and $Na_2IrO_3$ [26] with an effective spin $j_{eff} = 1/2$. Interestingly these systems are the most promising candidates to realize Kitaev QSL state, where the combined effect of Heisenberg-Kitaev model through strong spin-lattice coupling gives rise to exotic behaviors.

The co-existence of both prerequisites as mentioned above, for the realization of the Kitaev interactions makes NCTO potential candidate for further investigation of the Kitaev physics. In *3d* cobaltates, the $d^7$ $Co^{2+}$ ions in the octahedral crystal field with high spin configuration ($t_{2g}^5 e_g^2$; S=3/2 and L=1) forms a spin-orbit entangled pseudospin $j_{eff} =1/2$ (see fig. S1(c) for pictorial representation) [21,27]. The presence of $e_g$ spin contribution to J is ferromagnetic in nature and it largely compensates the antiferromagnetic contribution from $t_{2g}$ orbitals. This results in the reduction of J value and implies J<< K which may results into spin liquid ground state [10]. This suggest, NCTO is a potential spin-orbit entangled frustrated magnet and shows very rich physics in the low temperature regime [19]. The observation of a broad magnetic continuum in our measurement consistent with theoretical expectations suggest the possibility of the existence of the Kitaev QSL state in NCTO. For the current system, the long-range zigzag AFM ordering is observed at ~ 25-30 K [19,28–35], while a crossover temperature between quantum paramagnetic phase to a pure paramagnetic phase is reported around ~ 140-150

K [31,32]. Also spin reorientation phenomena is reported in the vicinity of ~ 17 K [28,29,31,33,35,36]. Additionally, non-centrosymmetric structure is a necessary condition for ferroelectricity and NCTO exhibit this non-centrosymmetricity and hence it may exhibit the multiferroic and/or magnetoelectric properties [29,36]. Interestingly a ferroelectric transition for NCTO is also reported in the temperature range of ~ 70-75 K [34]. The co-existence of the magnetic ordering as well as the ferroelectric ordering and other exotic properties such as Kitaev QSL state makes NCTO as a potential candidate for applications in quantum computation and device application. Hence, it is important to probe theses transitions and investigate and study quasi-particle excitations using temperature dependent Raman spectroscopic studies. Raman spectroscopic technique is a powerful technique to probe the spin, phonon, orbital, fractionalized excitations, and electronic excitations and study the phenomena associated with these excitations in 2D materials like transition metal dichalcogenides, quantum spin liquid candidates and other magnetic materials [34-44].

Hence, we performed a detailed temperature dependent Raman spectroscopic investigation of $Na_2Co_2TeO_6$, in the temperature range of ~ 5 to 300 K covering a broad spectral range from 2 to 2000 cm$^{-1}$ using 633 nm laser excitation, to understand the important phenomena in this 2D putative Kitaev QSL candidate. We observed signature of multiple phase transitions i.e. long-range zigzag AFM transition ($T_N$) at ~ 30 K, ferroelectric transition ($T_{FE}$) at ~ 70 K, and a crossover from pure paramagnetic phase to a quantum paramagnetic phase around $T^*$ ~ 150 K reflected in the renormalized self-energy parameters of the Raman active phonon modes. Interestingly, we observed a clear signature of spin reorientation within the AFM phase around $T_{SR}$ ~ 17 K. We also observed a asymmetric phonon mode in the low frequency region, and it appears below the transition temperature ~ 70 K, which potentially corresponds to a structural transition and/or underlying magnetic excitations [34,48]. Signature of the underlying broad magnetic continuum in the quantum paramagnetic phase and its temperature dependence suggest presence of frustrated magnetic interaction in the quantum paramagnetic phase. In the low frequency regime, quasi-elastic scattering (QES) shows the temperature dependence as expected for the low dimensional systems or frustrated magnets. To unravel symmetry of the phonon modes, a polarization dependent study at four different temperatures were performed. Our investigation of the Raman spectra as function of temperature reveals multiple phase transitions associated with spin-spin correlation (spin reorientation, long and short-range magnetic ordering), a potential polar order, and the temperature dependence of the broad magnetic continuum.

## 2. Experimental and computational details

### 2.1. Experimental details

Temperature dependent Raman scattering measurements were performed using the Lab RAM HR-Evolution Raman spectrometer with 1800 lines/mm grating in the backscattering configuration and closed-cycle He-flow cryostat (Montana) from 5 to 300 K with ± 0.1K accuracy. The spectra were excited using 633 nm laser excitation and the laser power was kept very low (< 0.5 mW) to avoid the local heating on the sample. A 50x (NA 0.8) long working distance objective was used to focus the laser light on the sample as well as to collect the scattered light from the sample. To unveil symmetry of the phonon modes, we have performed polarization dependent measurement using 633 nm laser excitation.

### 2.2. Computational details

NCTO belongs to the space group $P6_322$ (#182). Na occupies six lattice sites with a fractional occupancy of 67 %, which effectively gives rise to a total of four Na atoms in the unit cell. By choosing four positions of those six available ones, we get fifteen possible combinations of NCTO unit cell. By comparing the ground state energies of all these structures, we choose the minimum energy configuration NCTO. Here, the Co atoms with partially filled d-orbitals lead to the magnetic states of NCTO. We chose the AFM configurations without spin-orbit coupling and took the minimum energy configuration for further analysis after optimisation (see fig. S4 [27]). We performed phonon calculations at Γ-point for the configuration which has the lowest energy.

We used Vienna ab initio simulation package (VASP) [49,50] for the density functional theory (DFT) calculation within a plane-wave basis set. The generalized gradient approximation (GGA) of the revised Perdew-Burke-Ernzerhof for solids (PBEsol) [51,52] exchange correlation and projector augmented wave potentials [50,53] are used. The pseudopotentials have valence configurations $3s^1 3p^0$ for Na, $3d^8 4s^1$ for Co, $5s^2 5p^4$ for Te, and $2s^2 2p^4$ for O. The effective Hubbard interactions $U_{eff}$ [54] for magnetic Co atoms are treated in the Dudarev scheme [55]. The $U_{eff}$ for Co is chosen to be 5 eV [24]. We did the magnetic calculations without spin-orbit coupling (SOC). We set the kinetic energy cutoff of the plane wave basis as 500 eV and used a $7 \times 7 \times 3$ gamma centred K-mesh. We do the symmetry-protected volume and ionic relaxation by the conjugate-gradient algorithm until the Hellman-Feynman forces on each atom reach a value below the tolerance value of $10^{-3}$ eV/°A. We calculated the harmonic phonon modes at the gamma point of NCTO using the finite-displacement method as implemented in the VASP [49,50] code. We used an $11 \times 11 \times 5$ k-point mesh to calculate the forces.

## 3. Results and Discussions

### 3.1. Lattice Vibrations in $Na_2Co_2TeO_6$

Bulk single crystal of two-dimensional $Na_2Co_2TeO_6$ belongs to the non-centrosymmetric space group $P6_322$ (#182) and point group $D_6$ (222) in the hexagonal structure and it contained edge sharing $CoO_6$ (and $TeO_6$) octahedra that form a perfect honeycomb lattice of the $Co^{2+}$ ions [34,36]. According to the group theory, bulk single crystal of NCTO containing four formula unit consist of 44 atoms per unit cell which result into the 132 phonon branches at the Γ-point of the Brillion zone and it is expressed by the following irreducible representation as $\Gamma_{irred.} = 9A_1 + 13A_2 + 12B_1 + 10B_2 + 22E_1 + 22E_2$. There are 129 optical and three acoustic branches with $\Gamma_{optical} = 9A_1 + 12A_2 + 12B_1 + 10B_2 + 21E_1 + 22E_2$ and $\Gamma_{acuostic} = A_{2u} + E_{1u}$. Among the optical branches, $A_2$ is IR active, $A_1$ and $E_2$ are Raman active while $E_1$ is both Raman and IR active, $B_1$ and $B_2$ are Hyper-Raman active modes [56]. For the current system, The phonon frequencies obtained from the first-principles DFT calculations are listed in Table-I. Figure 1(a) shows the Raman spectrum of single crystal NCTO in the spectral range of ~ 200-720 cm$^{-1}$ recorded at the lowest temperature 5 K. Spectra were fitted using a sum of Lorentzian functions to extract the self-energy parameters of the phonon modes i.e. mode frequency ($\omega$) and full width at half maximum (FWHM) as well as the intensity. We observed more than twenty-seven Raman active phonon modes at 5 K, and the observed phonon modes with frequency are listed in Table-I. Figures 1(b) and 1(c) show the temperature evolution of the Raman spectrum in the frequency range of 30-615 cm$^{-1}$ and 620-710 cm$^{-1}$, respectively; demonstrating the softening and broadening of the peaks with rise in temperature. Furthermore, in the low frequency regime, an additional mode S* (~ 63 cm$^{-1}$) is observed with asymmetric line shape and surprisingly it appears only below ~ 50 K. We also observed a very weak phonon mode P* (~ 324 cm$^{-1}$; see fig. 1(b)) which appears below ~ 70 K. We further observed the quasi-elastic Raman scattering in the low frequency regime and very sharp peaks (probably due to air [57]) are also observed in the low frequency regime. Additionally, some weak modes attributed to the crystal field excitations (CFE) are observed marked by CFE1, CFE2, CFE3, and CFE4, see Fig.4 for their temperature evolution [48]. The peaks CFE1, CFE2, CFE3, and CFE4 appear below ~ 150 K as CFE becomes more prominent with lowering temperature as expected due to increase in the ground state population as govern by the Boltzmann distribution. In the high frequency region, we observed an underlying broad continuum which also shows temperature dependence as well as some weak modes which may be second-order phonon modes (marked as M1, M2, M3, and M4; see fig. 4 (c)) [27].

### 3.2. Temperature dependence of the phonon modes

Figures 2 (a-d) shows the temperature dependence of the frequency ($\omega$) and linewidth (FWHM) of the selected prominent phonon modes P4, P7, P8, P12, P14, P15, P20, P23, P24, and P25. We made the following important observations: (1) All the phonon modes show overall hardening, except P14 which shows hardening till ~ 150 K, with lowering temperature from 300 K to ~ 30 K with some change in slope at ~ 150 K, ~ 70 K, and ~ 30 K. Below ~30 K, frequency of the modes P4, P14, and P24 becomes nearly constant and other modes shows the downward trend (softening) suggesting the strong spin-phonon coupling in the long-range ordered AFM phase. We note that, apart from long-range ordered phase transition effect the mode frequency shows clear change around ~ 150 K (P4, P7, P14, P20 and the strongest mode P23) clearly suggesting emergence of possible transitions in this system. Quite interestingly, some of the modes do show changes in mode frequency and linewidth even within the AFM phase. For examples mode P7, P8, P15, P23 (the strongest mode), P24 and P25 show changes around ~ 15K, though changes are small. The frequency decreases with increase in temperature from ~ 150 -300 K for the phonon modes P4, P7, P8, P12, P14, P15, P20, P23, P24, and P25: and the change is ~ 3.3, 2.5, 2.9, 3.1, 4.9, 2.3, 0.7, 1.8, 2.7, and 3.0 cm$^{-1}$ ; respectively. (2) Linewidth of all the modes shows overall decreasing behavior with lowering the temperature till ~ 30 K ($T_N$); with a clear change in slope around ~ 150 K (P4, P7, P8, P12, P14, P15, P20, P24) and 70 K (P12, P20, P23, P24), similar to their frequency counterpart. Below ~ 30 K, linewidth of the modes P20 and P24 shows narrowing; on the other hand modes P8, P12, P15, and P25 shows broadening. Similar to the frequency, linewidth of some of the modes do shows changes deep inside the long-range ordered AFM phase i.e. around ~ 17 K. In particular, linewidth of the modes P7, P8, P12, P24, and P25 do shows change, though small. (3) In totality, we observed a clear changes in the frequency and linewidth of the phonon modes around ~ 150 K, 70 K, and 30 K as well as at ~17 K. The change in mode frequency and linewidth around ~ 30 K ($T_N$) and at ~ 17 K are attributed to the strong spin-phonon coupling as the system enters in the long-range ordered AFM phase and reorientation of the spins within the AFM phase, respectively. We note that linewidth of many modes shows linewidth broadening below ~ 30-40 K and it is opposite to the conventional anharmonic behaviour, where a linewidth is expected to decrease with decreasing the temperature. This opposite trend suggests the availability of additional decay channel, i.e., magnons, and as a result phonons will decay faster, which results in the increase in their linewidth (as lifetime (τ) α 1/FWHM). We note that changes observed below ~ 17 K were small, so in order to decipher this transition we did our measurement at a much smaller interval of temperature (~ 2 K) from 5 - 40 K. Figure 3(a) show the temperature dependence of frequency of few prominent phonon modes P23-P25 in the temperature range of 5-40 K. For modes P23-P25, frequency increases with increase in temperature up to $T_{SR}$ ( ~ 17 K), and decreases with further increase in temperature with a change in slope around $T_N$ (~ 30 K). This shows the clear signature of spin-reorientation around ~ 17 K.

With lowering the temperature, a structural transition ($T_{FE}$ ~ 70 K) is suggested from P6$_3$22 (#182) to polar P6$_3$ (#173) structure, which results into a polar order leading to the ferroelectric ordering associated with the distortion of CoO$_6$ octahedra [34]. We also observed that linewidth (for the modes P7, P8, P12, P23, and P24) and peak frequency (for the modes P4, P7, P12, P14, P15, P23, and P24) remains nearly constant from ~ $T_{FE}$ (~ 70 K) till ~ $T_N$ suggesting changes due to the polar ordering. A distinct change in the mode frequency and linewidth is also seen at $T^*$ (~ 150 K) indicating the transition from a pure paramagnetic phase to the quantum paramagnetic phase. Renormalization of the phonon modes around ~ 150 K suggest the coupling of phonons with the underlying magnetic excitations and this may be the reflection of the emergence of nontrivial spin excitations much above the long-range ordering temperature, similar to putative Kitaev - QSL candidate $\alpha$-RuCl$_3$. We note that a broad peak in the magnetic specific heat as well as a change in the susceptibility measurement were reported around same temperature for NCTO [31,32] and a two-step release of the magnetic entropy with only ~ 50% of the expected entropy is recovered across the magnetic transition, implying that nearly half of the entropy is released above $T_N$ signalling existence of a nontrivial spin excitations much above long range magnetic ordering temperature. Bera et al., [31] advocated that above ~150 K a paramagnetic state exists and with decreasing temperature below ~ 50 K an onset of short-range magnetic correlations is established and strong spin fluctuations down to 2 K. There observations also hints that there is nontrivial spin excitations in this system which starts building up below ~ 150 K. Our observation of phonon renormalization clearly suggest the emergence of nontrivial spin excitations below ~ 150 K, which may be responsible for the renormalization of the phonon modes. These anomalies needs to be further explored to look into the effect of potential QSL phase on the phonon modes in the quantum paramagnetic phase below ~ 150 K, similar to the case of K-QSL candidate $\alpha$-RuCl$_3$, where coupling of phonons with the Kitaev magnetic excitations is suggested [58,59].

The change in frequency of the phonon modes from ~ 150-330 K is mainly due to the anharmonic effect which arises from the phonon-phonon coupling. To understand temperature dependence of the first order phonon modes, we have fitted frequency and FWHM of the phonon modes in the temperature range of 150-300 K with an anharmonic model including three and four phonon decay channels given as [60]:

$$\omega(T) = \omega_0 + A\left(1 + \frac{2}{(e^x - 1)}\right) + B\left(1 + \frac{3}{(e^y - 1)} + \frac{3}{(e^y - 1)^2}\right), \text{ and} \tag{1}$$

$$\Gamma(T) = \Gamma_0 + C\left(1 + \frac{2}{(e^x - 1)}\right) + D\left(1 + \frac{3}{(e^y - 1)} + \frac{3}{(e^y - 1)^2}\right) \tag{2}$$

Where, $\omega_0$ and $\Gamma_0$ are the phonon mode frequency and linewidth at T= 0 K, respectively; $x = \frac{\hbar\omega_0}{2k_B T}$, $y = \frac{\hbar\omega_0}{3k_B T}$; A, B, C, and D are constant associated with change in frequency and FWHM as function of the temperature. The solid red lines in Figs. 2(a-d) are fitted curves with equations (1) and (2), the best fitted parameters are listed in Table-II. Temperature dependence of the phonons below 150 K is very crucial, clearly suggesting the emergence of multiple-phase transitions in the renormalized phonons self-energy parameters.

Interestingly, in the low frequency region, an asymmetric phonon mode, marked as S* (~ 63 cm$^{-1}$), is observed (see fig. 1(b)) and it appears only below ~ 50 K. We note that similar excitation observed around ~ 55 cm$^{-1}$ has been attributed to the magnetic excitations other than the magnon, which splits into two branches under an external magnetic field [48,61]. We have also attributed this mode to the magnetic excitation as well. The asymmetric line shape may be due to the interaction between the discrete quasi-particles (phonon here) and the underling continuum. To understand its asymmetric line shape, it is analysed using Fano profile given as [62]; $I(\omega) \propto \frac{(1 + \delta/q)^2}{(1 + \delta^2)}$, where $\delta = \frac{\omega - \omega_0}{\Gamma}$, and $I(\omega)$ is the Raman intensity which is a function of frequency $\omega$. $\omega_0$ and $\Gamma$ are the frequency and FWHM of the uncoupled phonon, respectively. q is the asymmetry parameter and the coupling strength between the phonon and the underlying continuum is quantifying by $1/|q|$. In the limit $1/|q| \to \infty$; coupling is stronger which results in the more asymmetric line shape, while in the limit $1/|q| \to 0$; coupling is weak and results into the Lorentzian line shape. Figure 3(b) shows the fitted Raman spectra for the mode $S^*$ with a Fano function at three different temperatures i.e. 15, 35, and 50 K. Figure 3(c) shows the dependence of the coupling strength $1/|q|$ as a function of temperature in the range 5-50 K. The changes in slope are observed around $T_{SR}$ (~ 17 K) and $T_N$ (~ 30 K) reflecting the effect of magnetic ordering on the Fano interaction. With increasing temperature $1/|q|$ value increases and shows the small jump around $T_{SR}$ ~ 17 K and become maximum around $T_N$ ~ 30 K and it decreases with further increasing the temperature.

The development of a magnetic ordering (short as well as long) and other phase transitions may potentially influence the Raman scattering cross section, which may be observed via evolution of the phonon modes intensity [63–65]. The ferroelectric phase transition may also influences the temperature dependent behaviour of the modes intensity. Figure 3(d) and 3(e) shows the temperature dependence of the intensity of the phonon modes P4, P5, P12, P18, P20, and P25 in the temperature range of 5-300 K. Intensity of the P4 mode decreases with increasing temperature up to ~ 30 K, and with further increase in temperature the change in the slope is observed around ~70 K and 150 K, respectively. For the modes P5, P12, P20, and P25 the intensity decreases with increase in temperature up to ~ 30 K, and then becomes nearly constant up to ~ 70 K and decreases rapidly with further increase in temperature with change in slope around ~ 150 K. Intensity of the mode P17 decreases with increasing temperature up to ~ 30 K and slightly increases with increase in temperature up to ~ 70 K, and it decreases significantly with increase in temperature with change in slope around ~ 150 K. The change in slopes around ~ 30 K ($T_N$), ~ 70 K ($T_{FE}$), and ~ 150 K ($T^*$) are associated with the zigzag AFM long-range ordering, ferroelectric transition, and crossover temperature from quantum paramagnetic to pure paramagnetic phase, respectively. We also observed small change in the slope for the phonon modes P12 and P20 around $T_{SR}$ (~17 K) (see insets of figs. 3(d) and (e).

### 3.3. Crystal field excitations and Kramer's doublet

Crystal field excitations appears due to the lifting of the orbital degeneracy of transition metal ions in electrostatic field due to surrounding ions. Investigation of the crystal field excitation and their coupling with phonons may provide crucial understanding about the thermal, electronic and magnetic as well as superconducting properties of the various magnetic and other such materials [66]. In our measurements, we observed weak modes around ~ 22.9 meV (~ 185 cm$^{-1}$), 23.3 meV (~ 188 cm$^{-1}$), 119 meV (~ 967 cm$^{-1}$), 137.5 meV (~ 1110 cm$^{-1}$), and 152 meV (~1223 cm$^{-1}$) at 5 K. We note that these modes are intense only at low temperature, reflecting their nature as crystal field excitations as per Boltzmann distribution (because the ground state population increases with decreasing temperature and hence their intensity). Recently, Mou et al., [67] also reported crystal field excitations in this system at ~ 22 meV (~ 177 cm$^{-1}$), ~ 69 meV (~557 cm$^{-1}$), 118 meV (~ 1015 cm$^{-1}$), and ~138 meV (~ 1186 cm-1) which are attributed to the octahedral crystal field splitting of Co$^{2+}$ ions level. We observed a mode around ~ 69 meV, but with a distinct behaviour as compared to other phonon modes suggesting its overlapping/strong coupling with underlying CFE. In Fig. 4 (a), for the phonon mode P18 (~ 69 meV), frequency increases with increase in temperature up to ~ 100 K, and with further increase in temperature it remains nearly constant till ~ 150 K and then decreases afterwards, suggesting a strong coupling between CFE and the phonon up to ~ 150 K [68]. The integrated intensity of the mode P18 decreases gradually with increase in temperature which is typical for Raman scattering associated with CFEs [69]. Mode P18 shows the nearly two-fold symmetry at 5 and 50K, while at 100 and 300K it shows a quasi-isotropic behaviour. This change in symmetry with change in temperature may results due to the strong coupling between CFE and phonon at low temperatures. Figure 4(b) and 4(c) shows the temperature evolution of the Raman peaks which are attributed to the crystal field excitations marked as CFE1 (~185 cm$^{-1}$), CFE2 (~ 967 cm$^{-1}$), CFE3 (~ 1110 cm$^{-1}$), and CFE4 (~1223 cm$^{-1}$). The CFE1 consist of two

Raman peaks R1 and R2 among which R1 appears below ~120-130 K and the separation of these two peaks is ~ 3 cm$^{-1}$ (~ 0.4 meV) at 5 K. This suggest the lifting of Kramer's degeneracy of Co$^{2+}$ ion at lower temperature [68].

### 3.4. Temperature dependence of the underlying magnetic continuum

The presence of various magnetic excitations and underlying magnetic degrees of freedom help to shed light on the possible exotic phases such as QSL and other important properties, i.e., phase transitions with various external perturbations (i.e., temperature, pressure, etc.) in low-dimensional materials. To understand the temperature dependence of the observed magnetic excitations and the underlying physics associated with dynamics of these magnetic excitations, we study and analyzed the Raman spectrum as a function of temperature in the spectral range of 3-1400 cm$^{-1}$. In the low frequency region, we observed a quasi-elastic Raman scattering; while in the high frequency region, a broad continuum is observed, and both of these shows temperature dependence.

In magnetic materials, fluctuations of the magnetic energy density as well as spin fluctuations result in the quasi-elastic Raman scattering and it contributes to magnetic specific heat (C$_m$). The magnetic specific heat provides an important insight for the underlying magnetic excitations, spin energy fluctuations, and the Raman signature of the phase transitions associated with a change in temperature [70–72]. We extracted the dynamic Raman susceptibility, $\chi^{dyn}$, associated with the observed low frequency quasi-elastic scattering. The Raman response, $\chi''(\omega)$, is an imaginary part of the susceptibility and reflects the dynamic properties of the underlying collective excitations. It is obtained from the raw Raman intensity simply dividing by the Bose thermal factor $\left(I(\omega) \propto \chi''(\omega)[n(\omega)+1]\right)$. Figure 5(a) shows the temperature evolution of the Raman response in the spectral range of ~ 3-75 cm$^{-1}$. In the high temperature regime, Raman response is weak and with lowering temperature, it exhibits strong enhancement. This reflects enhanced fluctuations in the low temperature regime as compared to high temperature regime. To extract dynamic Raman susceptibility $\chi^{dyn}$, we first determine the Raman conductivity from Raman response divided by Raman shift $\chi''(\omega)/\omega$. Figure 5(b) shows the temperature evolution of the Raman conductivity. $\chi^{dyn}$ is extracted using Raman conductivity and given by the Kramers-Kroning relation:

$$\chi^{dyn} = \lim_{\omega \to 0} \chi(k=0,\omega) \equiv \frac{2}{\pi} \int_0^\Omega \frac{\chi''(\omega)}{\omega} d\omega, \qquad (3)$$

i.e. by extrapolating data from lowest measured frequency to 0 cm$^{-1}$. The upper cutoff frequency ($\Omega$) is taken as 75 cm$^{-1}$. Figure 5(c) shows the temperature dependence of the $\chi^{dyn}$. We fitted $\chi^{dyn}$ with a power law as $\chi^{dyn} \propto T^\beta$ (solid red line; $\beta = -0.38$). In the hydrodynamic limit [73], the magnetic specific heat ($C_m$) can be extracted from the quasi-elastic scattering part and the Raman conductivity and magnetic specific heat are related as [70]:

$$\frac{\chi''(\omega)}{\omega} \propto C_m T \frac{Dk^2}{\omega^2 + (Dk^2)^2} \qquad (4)$$

Where, D is the thermal diffusion constant ($D = k/C_m$) and $k$ is the thermal conductivity. Fig. 5(d) shows temperature dependence of the magnetic specific heat $C_m$. With increasing temperature, $C_m$ decreases with changes around $T_{SR}$ ~ 17 K, $T_N$ ~ 30 K, $T_{FE}$ ~ 70 K, and $T^*$ ~ 150 K, and agree with phase transitions observed via renormalization of the self-energy parameters of the phonons.

In the high frequency region, we observed a broad continuum in the spectral range of ~ 330 to 1400 cm$^{-1}$ (see fig. 5(e)), which shows its strong temperature dependence. With increasing temperature, its intensity remains nearly constant up to $T_N$ (~ 30 K) with a jump/drop around this temperature and decreases slightly till $T_{FE}$ (~ 70 K) and with further increase in temperature it decreases abruptly with a slope change observed around $T^* \sim 150$ K. The experimental measurements like Inelastic Neutron Scattering (INS) [61,74], Nuclear Magnetic Resonance (NMR) [75], and transport measurements [29,31,32] as well as theoretical studies [10,20,21,76] on NCTO suggest a proximity towards QSL ground state. In our measurements, we observed high frequency underlying broad continuum which shows the significant temperature dependence. To shed more light on the behavior of this underlying magnetic continuum, we call for the further experimental and theoretical investigation for the current and other such more systems.

### 3.5. Polarization dependence of the phonons and the underlying broad continuum

Polarization dependent Raman measurement provides an important insight for the symmetry of the lattice vibrations and dynamics of other quasi-particles excitation. To investigate and understand the dependence of symmetry of the phonon modes and the underlying continuum on the polarization and temperature, we performed polarization dependent measurement at four different temperatures i.e. 5, 50, 100, and 300 K by rotating the incident light systematically and scattering light polarization was kept constant.

Figure 6(a) shows the polarization dependent intensity of the phonon modes P12, P18, P20, and P23 at four different temperatures (see, Fig. S2 in the supplementary material for the polarization dependent evolution of the Raman spectrum at four different temperatures [27]). To understand the polarization dependence of the phonon modes intensity, within the semi-classical approach, intensity of the Raman active phonon mode is given as $I_{Raman} \propto |e_s.R.e_i|^2$, where T is the transpose, $e_i$ and $e_s$ are unit vectors in direction of incident and scattered light electric field. R is a Raman tensor for the phonon modes with corresponding symmetry. In matrix form, the above unit vectors can be written as : $\hat{e}_i = [\cos(\theta+\theta_0) \quad \sin(\theta+\theta_0) \quad 0]$, $\hat{e}_s = [\cos(\theta_0) \quad \sin(\theta_0) \quad 0]$, where $\theta$ is the relative angle between $\hat{e}_s$ and $\hat{e}_i$. $\theta_0$ is the angle of scattered light polarization from x axis as shown in the Fig. S1(d) [27]. For the current system, the Raman tensors for the phonon modes with symmetry $A_{1g}$, $E_{1g}$, and $E_{2g}$ are listed in the Table-III [77]. For the phonon modes with above mentioned symmetries, intensity depends on the polarization angle $(\theta)$, may be given as:

$$I_{A_{1g}} = |a[\cos(\theta_0)\cos(\theta+\theta_0) + \sin(\theta_0)\sin(\theta+\theta_0)]|^2 = a^2\cos^2(\theta), \quad 5(a)$$

$$I_{E_{1g}} = 0, \text{ and} \quad 5(b)$$

$$I_{E_{2g}} = |d[\cos(\theta_0)\cos(\theta+\theta_0) - \sin(\theta_0)\sin(\theta+\theta_0)]|^2, \text{ and} \quad 6(a)$$

$$I_{E_{2g}} = |d[-\cos(\theta_0)\sin(\theta+\theta_0) - \sin(\theta_0)\cos(\theta+\theta_0)]|^2 \quad 6(b)$$

Where, $\theta_0$ is an arbitrary angle. Without any loss of generality $\theta_0$ is taken as zero and hence equations 6(a) and 6(b) becomes:

$$I_{E_{2g}} = |d\cos(\theta)|^2, \text{ and} \quad 7(a)$$

$$I_{E_{2g}} = |-d\sin(\theta)|^2 \quad 7(b)$$

In Fig. 6(a) the solid red lines are the fitted curves using above equations. Phonon mode P12 (P4, P5, and P7 in supplementary figure S3) shows the circular symmetry at all temperatures. These modes are fitted with combination of equations 7(a) and 7(b). Modes P20 and P23 (P24 and P25 in supplemental Fig S3 [27]) shows the two-fold symmetry and are fitted using equation 5(a). Figure 6(b) shows the polarization dependence of the underlying broad continuum in the higher frequency regime, at four different temperatures, reflecting $A_g$ symmetry with maxima around $\sim 0^0$ and $\sim 180^0$.

Figure 7 shows the polarization dependence of the mode P16 which shows a two-fold symmetry with maximum intensity at $\sim 0^0$ and $\sim 180^0$ at 5 K. Surprisingly, with increase in temperature, main axis shows the maxima at $\sim 40^0$ and $\sim 220^0$ (rotated by $\sim 40^0$) at 50 K.

With further increase in temperature, it shows the maxima at $\sim 60^0$ and $\sim 240^0$ at 100 K. At 300 K, main axis shows the maxima at $\sim 90^0$ and $\sim 270^0$. The rotation observed for the mode P16 with increasing temperature is significantly large (about $\sim 90^0$) and this may be attributed due to the change in underlying phases. We note that changes are significant till ~ 100 K with almost linear increase in rotation angle of maxima ($\Psi$) and after that it increase slightly. A significant rotation of the intensity polar plot for mode P16 suggests that the Raman optical sectional rules are tunable by switching between the underlying phases as temperature changes [78]. In 2D magnetic materials, excitonic effect dominate the optical as well as magneto-optical responses which results in the strong magneto-optical Kerr effect [79]. The magneto optical effect is caused by the coupling of spin and charge degrees of freedom in crystals, and it has applications in the fields of detection, data storage, and optical modulation [78]. Hence, the quantitative analysis of the rotation of the intensity polar plot of the phonon mode P16 calls for a microscopic theory.

**Conclusion**

In conclusion, we performed temperature as well as polarization dependent Raman spectroscopic measurements on a single crystal of quasi 2D antiferromagnetic $Na_2Co_2TeO_6$. Our temperature dependent investigation gives a clear and distinct Raman signature of multiple phase transitions, magnetic and crystal field excitations, and a broad underlying magnetic continuum suggesting the possible origin in the fractionalized excitations. The polarization dependent measurements demonstrate a clear rotation of the major axis for one of the phonon mode; additionally exploring the mechanism of the underlying broad continuum in the high frequency range calls for a further theoretical investigation. Our results points that NCTO is a potential candidate for the realization of Kitaev physics. We expect our results to motivate new directions in Kitaev materials, especially focusing on the role of phonons and spin degrees of freedom and their intertwining.

**Acknowledgements**

P.K. thanks SERB (Project no. CRG/2023/002069) for the financial support and IIT Mandi for the experimental facilities.

**Data availabilty statement**

All data that support the findings of this study are included within the article and supplementary file.

**Figures:**

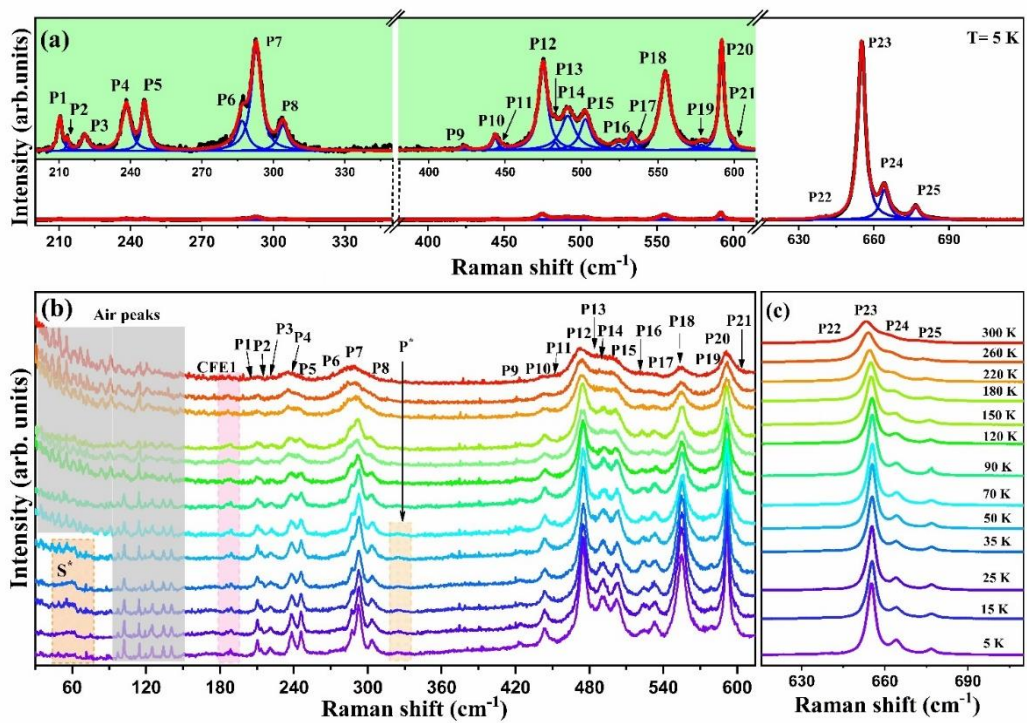

**Figure 1. (a)** Raman spectrum of Na$_2$Co$_2$TeO$_6$ in the spectral range of ~200 to 720 cm$^{-1}$ recorded at 5 K using 633 nm laser excitation. Insets in the green shaded area are the amplified spectra in the spectral range of 200 to 350 cm$^{-1}$ (left side) and 380 to 615 cm$^{-1}$ (right side). **(b)** Shows the temperature evolution of the Raman spectrum in the spectral range of 30 to 615 cm$^{-1}$. Shaded region around ~ 120 cm$^{-1}$ and low frequency at higher temperature shows air peaks. CFE1 is the crystal filed exist, S$^*$ and P$^*$ are the weak modes which appears below ~ 50 K and 70 K, respectively. **(c)** Shows the temperature evolution of the most prominent peaks P23, P24, and P25 in the Raman spectral range of 615 to 720 cm$^{-1}$.

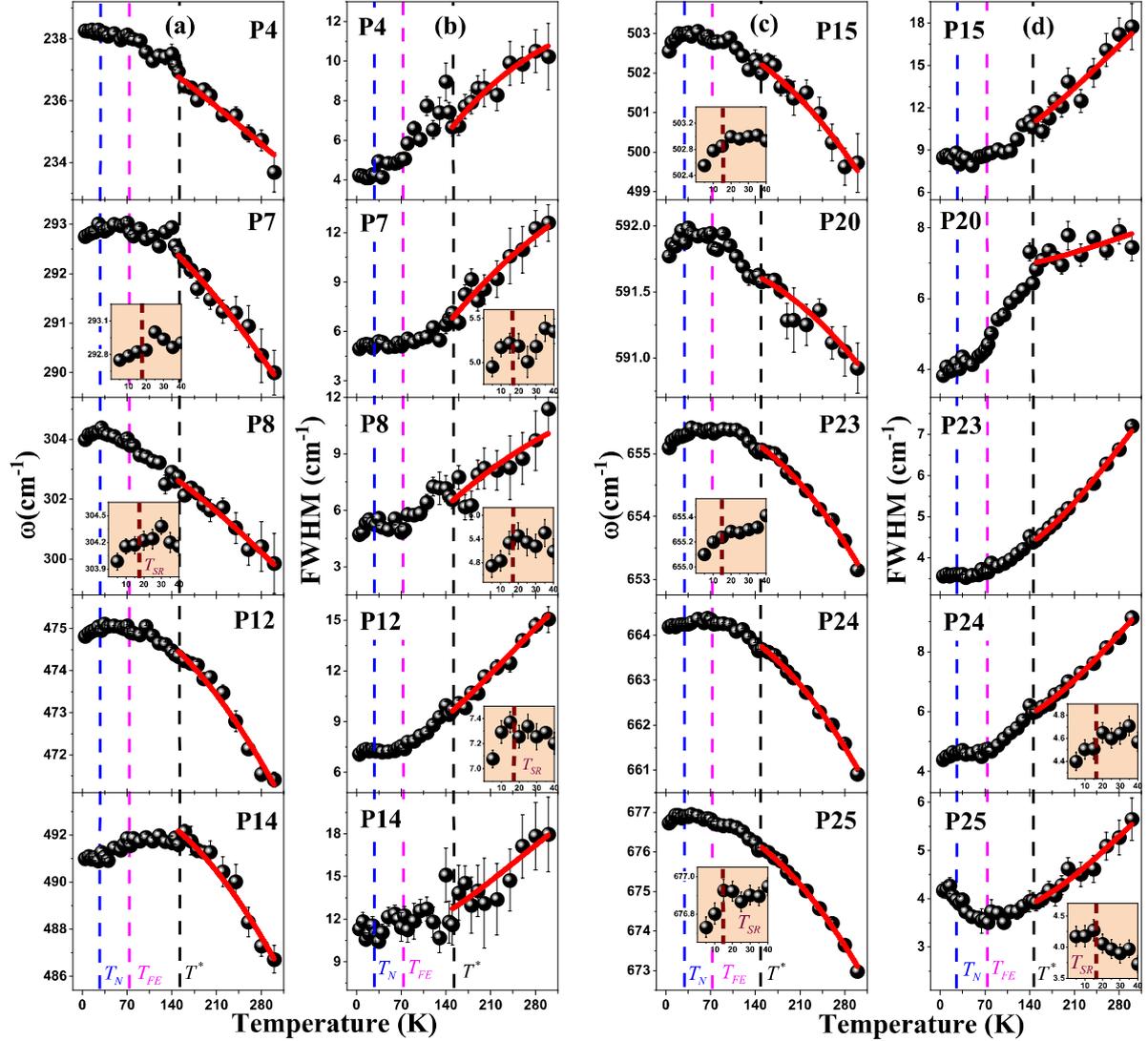

**Figure 2. (a)** and **(b)** Shows the temperature dependence of the frequency and FWHM of the phonon modes P4, P7, P8, P12, and P14; respectively. **(c)** and **(d)** Shows the temperature dependence of the frequency and FWHM of the phonon modes P15, P20, P23, P24, and P25; respectively (the solid red line shows a three and four phonon fitting in the temperature range of ~ 150 to 330 K). $T_N$ ~ 30 K, $T_{FE}$ ~70 K, and $T^*$ ~150 K represents the Zigzag AFM long-range ordering temperature, ferroelectric transition temperature, and crossover temperature, respectively. The inset shows the changes around spin reorientation temperature $T_{SR}$ ~ 17 K.

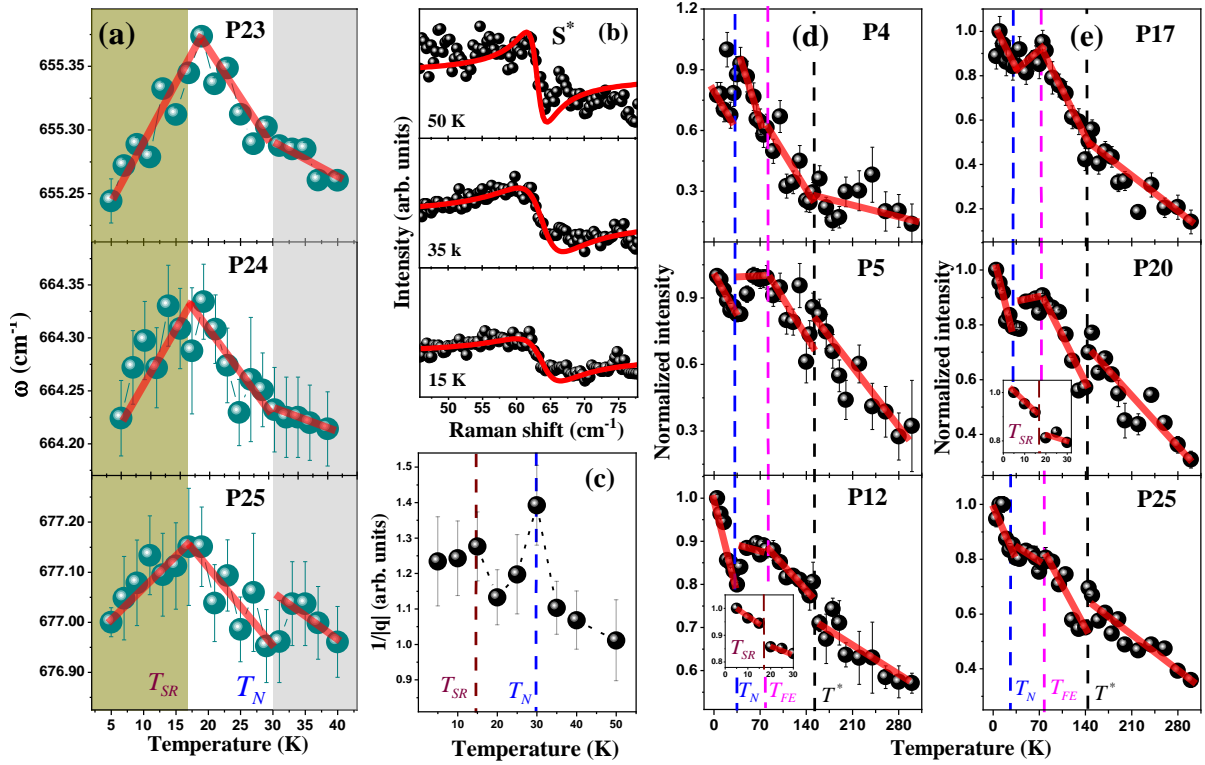

**Figure 3:** **(a)** Temperature dependence of the frequency of modes P23-P25 in the range of ~ 5-40 K, measured at an interval of ~ 2 K. **(b)** Shows the temperature evolution of the Fano peak S* at three different temperatures, plotted on the same intensity scale. **(c)** Temperature dependence of the asymmetry, $1/|q|$, in the temperature range of ~ 5-50 K. **(d)** and **(e)** Temperature dependence of the intensity of the phonon modes P4, P5, P12 and P17, P20, P25; respectively. Semi-transparent red lines are drawn as guide to the eye. Vertical dotted lines show the observed transition temperatures.

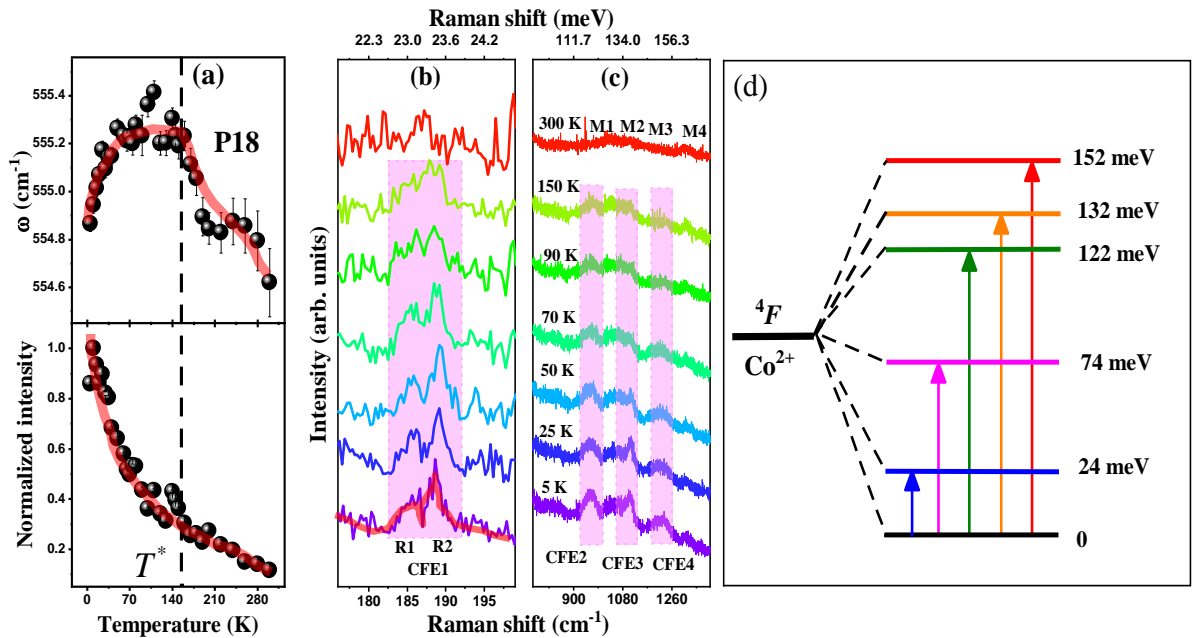

**Figure 4: (a)** Temprature dependance of the mode frequency and intensity of the mode P18. **(b)** and **(c)** Shows the temprature evolution of the crystal field excitations (CFE1, CFE2, CFE3, and CFE4) in the temprature range of 5-300 K, repectively. Semi-transparent red lines are drwan as guide to the eye. (d) Schematic for the energy levels of the ground $^4F$ ($Co^{2+}$) split by the crystal field

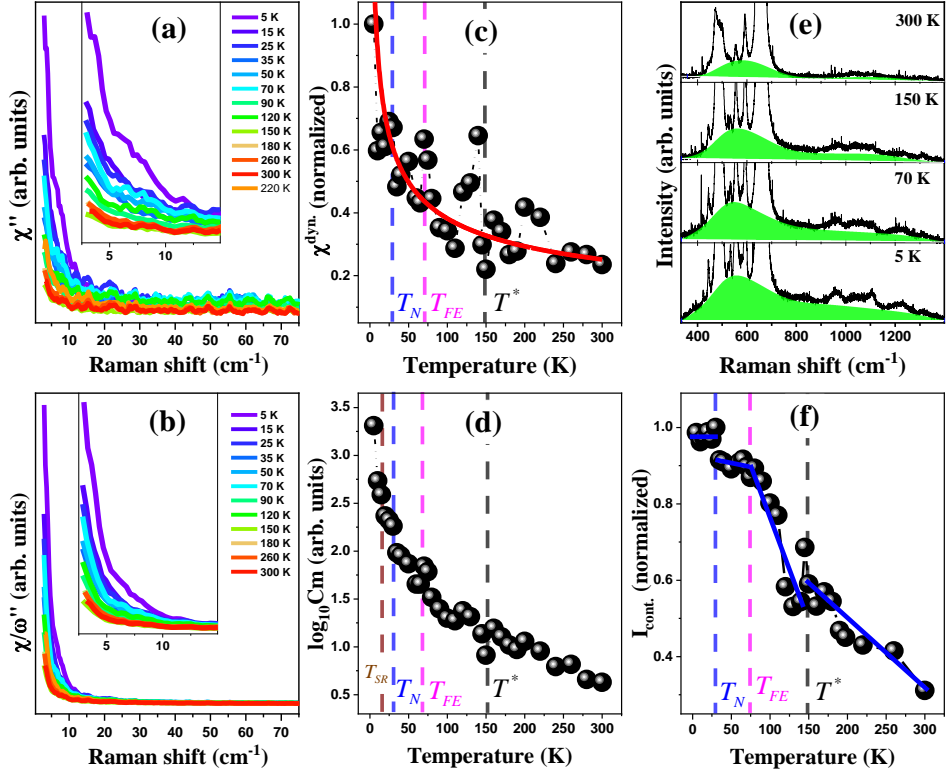

**Figure 5: (a)** and **(b)** Temperature evolution of the Raman response, $\{\chi''(\omega)\}$ and **(b)** Raman conductivity, $\{\chi''(\omega)/\omega\}$, in the temperature range of 5-300 K, respectively. **(c)** Temperature dependence of the dynamic Raman susceptibility $\chi^{dyn}$. The solid red line shows the fitted curve with power law $\chi^{dyn}(T) \propto T^\beta$ ($\beta = -0.38$). **(d)** Temperature dependence of the magnetic specific heat $C_m$ in the temperature range of ~ 5-300 K. **(e)** Shows the temperature evolution of the underlying broad continuum (green shaded area) in the spectral range of ~ 330-1400 cm$^{-1}$. **(f)** Shows the temperature dependence of intensity of the underlying broad continuum. The solid blue line is guide to the eye.

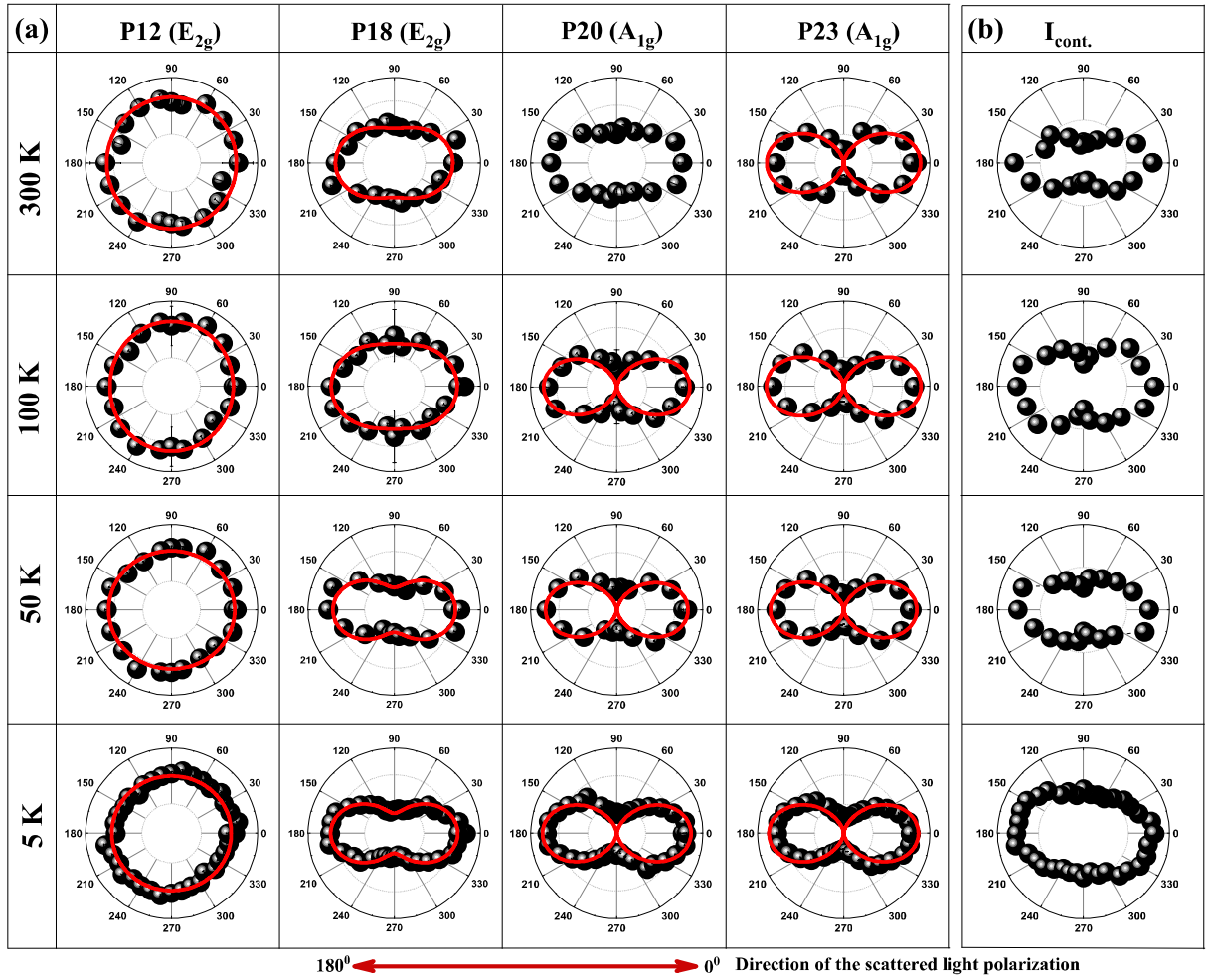

**Figure 6: (a)** Shows the polarization dependent intensity of the selected phonon modes P12, P18, P20, and P23 at four different temperatures 5, 50, 100, and 300 K. **(b)** Polarization dependence of the underlying broad continuum in the high frequency region, at four different temperatures. At the bottom, solid red line shows the direction of the scattered light polarization.

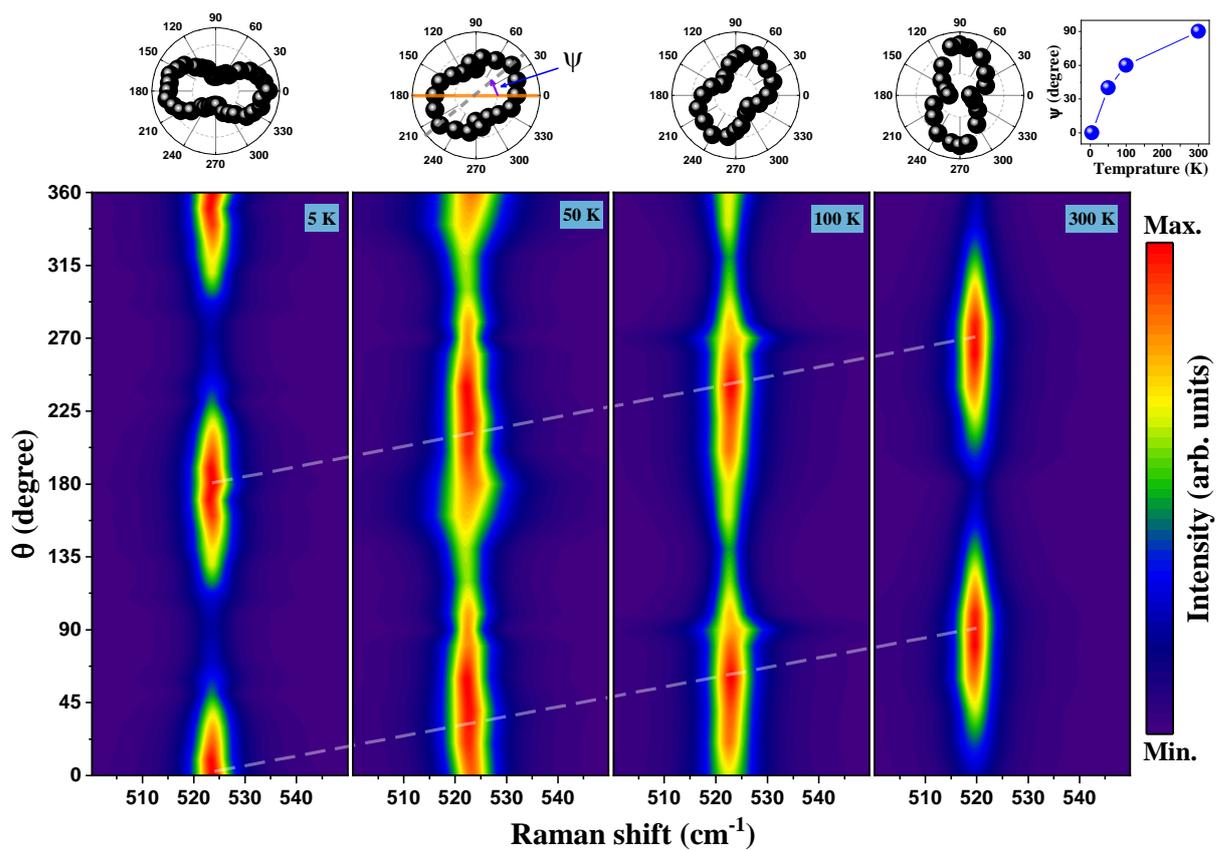

**Figure 7:** Upper panel shows the polarization dependent intensity polar plot and lower panel shows 2D colour contour map for the intensity the phonon mode P16 as a function of the polarization angle at four different temperatures 5, 50, 100, and 300 K.

**Table-I:** Experimentally observed and DFT-calculated Raman active phonon mode freqencies for single crystal NCTO. The units are in cm$^{-1}$.

| Raman peaks | Phonon freqency at 5 K (ω) | DFT-calculated phonon freqency | Raman peaks | Phonon freqency at 5 K(ω) | DFT-calculated phonon freqency |
|---|---|---|---|---|---|
| S* | 63.2 | 65 | P13 | 482.7 | - |
| P1 | 210.3 | 209 | P14 | 490.8 | - |
| P2 | 213.5 | 212 | P15 | 502.5 | - |
| P3 | 220.8 | 219 | P16 | 524.2 | - |
| P4 | 238.2 | 236 | P17 | 532.7 | - |
| P5 | 245.9 | 252 | P18 | 554.9 | - |
| P6 | 286.8 | 287 | P19 | 578.5 | 581 |
| P7 | 292.7 | 292 | P20 | 591.7 | 589 |
| P8 | 303.9 | 295 | P21 | 599.1 | 595 |
| P* | 324.2 | - | P22 | 638.6 | - |
| P9 | 423.7 | - | P23 | 655.1 | - |
| P10 | 443.7 | - | P24 | 664.2 | - |
| P11 | 446.5 | - | P25 | 676.2 | - |
| P12 | 474.8 | - | | | |

**Table-II.** List of the fitting parameters corresponding to the phonon modes in Na$_2$Co$_2$TeO$_6$, fitted using the **three and Four-phonon** fitting model in the temperature range of ~ 150 to 300 K. The units are in cm$^{-1}$.

| Mode assignment | $\omega_0$ (fixed) | A | B | $\Gamma_0$ (fixed) | C | D |
|---|---|---|---|---|---|---|
| P4  | 239.5 | -1.35±0.12 | -0.02±0.03 | 4.2 | -2.78±0.99 | 0.93±0.21 |
| P7  | 294.5 | -1.03±0.12 | -0.11±0.04 | 4.2 | -2.42±1.13 | 1.05±0.26 |
| P8  | 305.5 | -1.59±0.20 | -0.09±0.07 | 5.2 | -1.92±1.24 | 0.73±0.29 |
| P12 | 476.8 | -0.75±0.16 | -0.69±0.07 | 7.0 | -0.36±0.21 | 1.06±0.06 |
| P14 | 494.5 | 0.83±0.41  | -1.68±0.19 | 9.0 | 1.26±1.07  | 0.81±0.33 |
| P15 | 503.8 | -0.10±0.30 | -0.79±0.15 | 6.8 | 1.30±0.84  | 1.03±0.28 |
| P20 | 591.8 | 0.29±0.09  | -0.32±0.05 | 6.0 | 0.72±0.22  | 0.11±0.09 |
| P23 | 656.0 | 0.61±0.08  | -1.05±0.05 | 3.0 | 0.18±0.04  | 0.71±0.02 |
| P24 | 665.0 | 1.01±0.08  | -1.61±0.05 | 3.5 | 1.19±0.08  | 0.72±0.03 |
| P25 | 677.5 | 1.20±0.08  | -1.87±0.04 | 2.8 | 0.40±0.12  | 0.42±0.05 |

**Table-III:** Wyckoff positions and irreducible representations of phonon modes for honeycomb Na$_2$Co$_2$TeO$_6$ in the hexagonal (space group: P6$_3$22(#182)) phase.

| Atom | Wyckoff position | $\Gamma$ – point mode decomposition | Raman tensor |
|---|---|---|---|
| Co(1) | 2b | $A_2 + B_1 + E_1 + E_2$ | $A_{1g} = \begin{pmatrix} a & 0 & 0 \\ 0 & a & 0 \\ 0 & 0 & b \end{pmatrix}$ |
| Co(2) | 2d | $A_2 + B_1 + E_1 + E_2$ | |
| Te    | 2c | $A_2 + B_1 + E_1 + E_2$ | |
| O     | 12i | $3A_1 + 3A_2 + 3B_1 + 3B_2 + 6E_2 + 6E_1$ | $E_{1g} = \begin{pmatrix} 0 & 0 & 0 \\ 0 & 0 & c \\ 0 & c & 0 \end{pmatrix}, \begin{pmatrix} 0 & 0 & -c \\ 0 & 0 & 0 \\ -c & 0 & 0 \end{pmatrix}$ |
| Na(1) | 12i | $3A_1 + 3A_2 + 3B_1 + 3B_2 + 6E_2 + 6E_1$ | |
| Na(2) | 2a | $A_2 + B_2 + E_1 + E_2$ | $E_{2g} = \begin{pmatrix} d & 0 & 0 \\ 0 & -d & 0 \\ 0 & 0 & 0 \end{pmatrix}, \begin{pmatrix} 0 & -d & 0 \\ -d & 0 & 0 \\ 0 & 0 & 0 \end{pmatrix}$ |
| Na(3) | 12i | $3A_1 + 3A_2 + 3B_1 + 3B_2 + 6E_2 + 6E_1$ | |
| $\Gamma_{irred.} = 9A_1 + 13A_2 + 12B_1 + 10B_2 + 22E_1 + 22E_2$; $\Gamma_{optical} = 9A_1 + 12A_2 + 12B_1 + 10B_2 + 21E_1 + 22E_2$; and $\Gamma_{acoustic} = A_{2u} + E_{1u}$ | | | |

**Supplementary:**

# Raman signature of multiple phase transitions and quasi-particle excitations in putative Kitaev spin liquid candidate Na$_2$Co$_2$TeO$_6$

Atul G. Chakkar[1,*], Chaitanya B. Auti[1], Deepu Kumar[1], Nirmalya Jana[2], Koushik Pal[2], and Pradeep Kumar[1,#]

[1] School of Physical Sciences, Indian Institute of Technology Mandi, Mandi-175005, India

[2] Department of Physics, Indian Institute of Technology Kanpur, Kanpur-208016, India

*E-mail: atulchakkar16@gmail.com
#E-mail: pkumar@iitmandi.ac.in

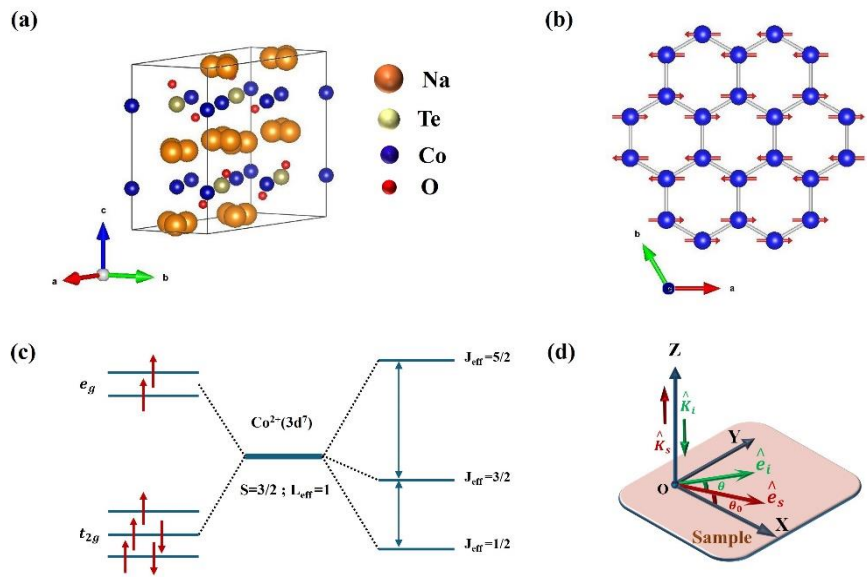

**Figure S1: (a)** Shows the crystal structure along an arbitary direction. **(b)** Shows the schematic for the layered honeycomb structure in the *ab*-plane with zigzag antiferromagnetic order. **(c)** Shows the schematic of splitting of the denegerate $d^7$ state in high spin configuration in the presence of an octahedral crystal field. **(d)** Schematic for the unit vectors ($e_i$ and $e_s$) of the incident and scattered light. $K_i$ and $K_s$ shows the direction of propagation of the incident and scattered light.

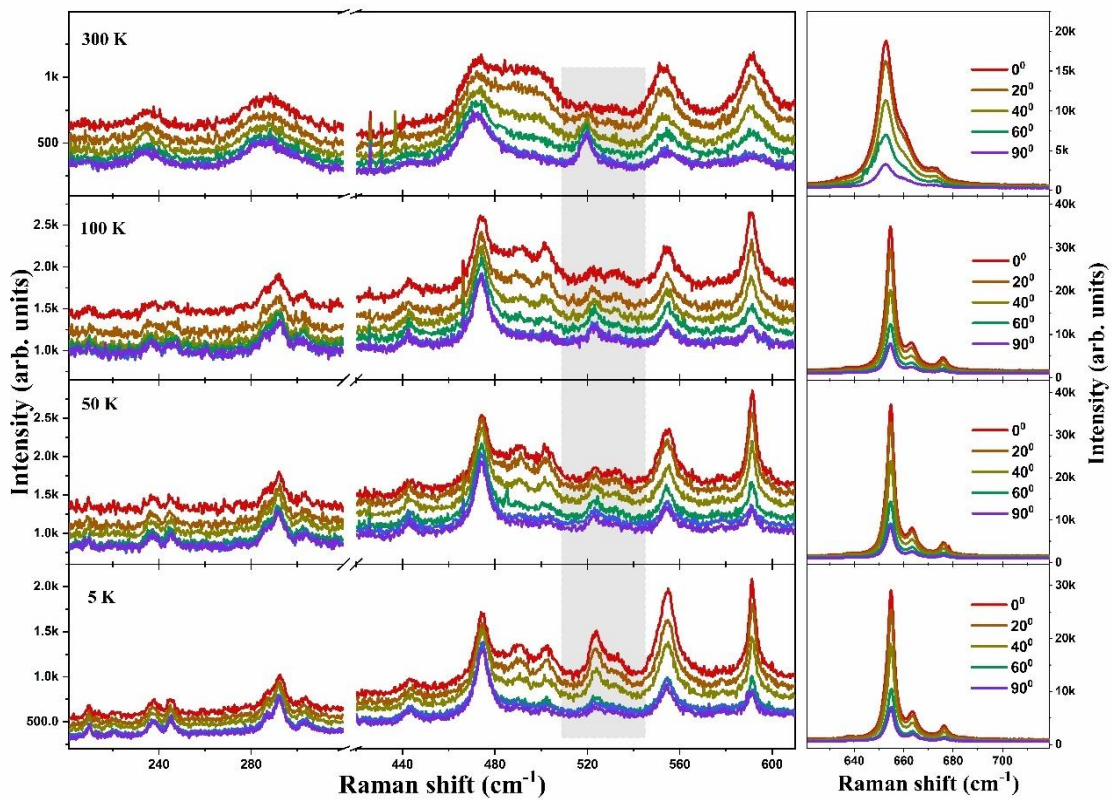

**Figure S2:** Shows the polarization dependence of the Raman spectrum at four different temperatures 5, 50, 100, and 300 K with incident light rotated configuration. The shaded grey area shows the polarization dependence of the phonon mode P16.

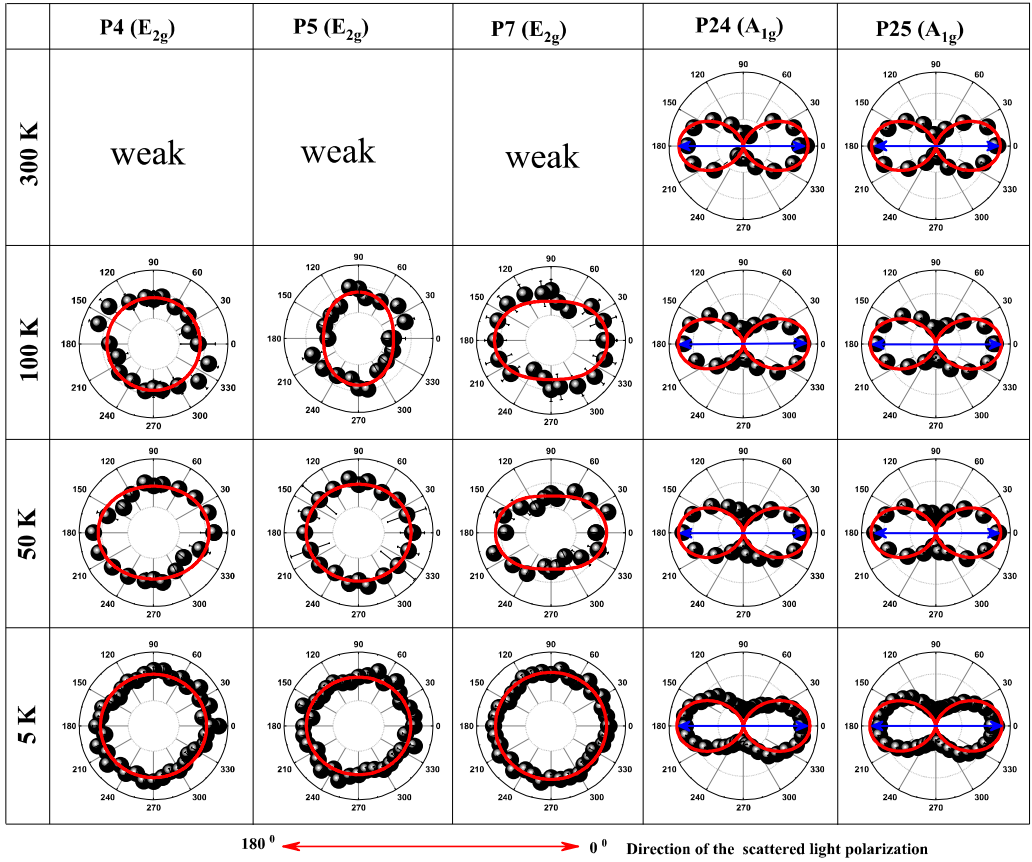

**Figure S3:** Polarization dependent intensity of phonon modes P4, P5, P7, P24, and P25 at different temperatures 5, 50, 100, and 300 K, respectively with incident light rotated configuration. At the bottom, solid red line shows the direction of the scattered light polarization.

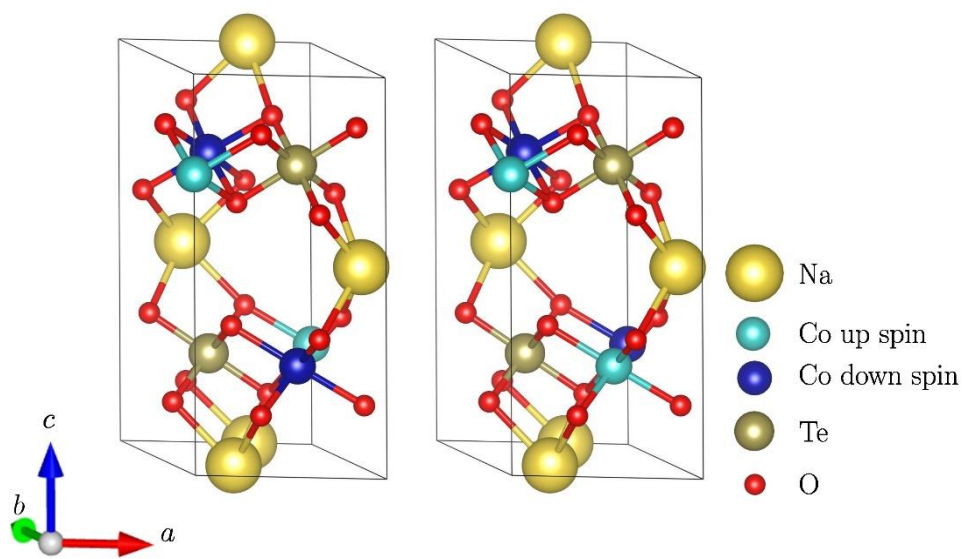

**Figure S4:** Two possible antiferromagnetic configurations of NCTO. The magnetic configuration of the Co atom is highlighted with cyan for spin-up and blue for spin-down state. The second configuration has the lowest energy.